\documentstyle[prl,aps]{revtex}
\begin{document}

\input epsf
\draft
\twocolumn[\hsize\textwidth\columnwidth\hsize\csname
@twocolumnfalse\endcsname                             
\title{Metastability and paramagnetism in superconducting mesoscopic disks}
\author{J. J. Palacios}
\address{Dept. de F\'{\i}sica Te\'orica de la Materia Condensada, 
Universidad Aut\'onoma de Madrid, Cantoblanco, Madrid 28049, Spain.}

\date{\today}
\maketitle

\widetext
\begin{abstract}
\leftskip 2cm
\rightskip 2cm

A projected order parameter is used to calculate, 
not only local minima of the Ginzburg-Landau energy functional, 
but also saddle points or energy barriers 
responsible for the metastabilities observed in superconducting 
mesoscopic disks (Geim et al. Nature {\bf 396}, 144 (1998)). 
We calculate the local minima 
magnetization and find the energetic instability points between vortex
configurations with different vorticity.
We also find that, for any vorticity, the supercurrent can reverse its 
flow direction on decreasing the magnetic field before one vortex can escape. 

\end{abstract}

\pacs{\leftskip 2cm PACS numbers: 74, 74.60.Ec, 74.76.-w}
\vskip2pc]

\narrowtext
The interest in understanding the creation and annihilation mechanisms and,
in general, the stability of vortices in superfluids has been recently 
boosted by a series of technological advances in both mesoscopic
superconductors\cite{Moshchalkov:95,Geim:97,Geim:98,Pardo:99}
and atomic condensates\cite{Butts:99}. 
Most of the proposals for the creation of 
vortices in atomic condensates face, at the present time,
severe technological difficulties. On the contrary, mesoscopic superconductors 
in magnetic fields are already proving to be
an ideal scenario where the detection and even manipulation 
of vortices at the individual level
is becoming more and more
feasible\cite{Moshchalkov:95,Geim:97,Geim:98,Pardo:99}.
A good example, although not the only one,
of single-vortex sensitivity can be found in 
the Hall magnetometry measurements performed on mesoscopic Al disks
by Geim et al.\cite{Geim:97,Geim:98}. Both field-cooled (FC) and 
constant temperature (CT) magnetization measurements  provide
evidence of the quantization of the vorticity of the order parameter.
When the system is kept out of equilibrium, it can show paramagnetic
response both in the CT and FC cases, 
whereas, as expected, equilibrium measurements {\em always} exhibit
diamagnetism. 

Geim's experiment sheds light on the controversial paramagnetic Meissner      
effect (PME) confirming that the PME, at least in mesoscopic samples,
is linked to a non-equilibrium magnetic response of the system. In
addition to this, it raises a number of fundamental questions concerning the  
metastability of few-vortex states in mesoscopic
systems: (i) What is the origin of the
metastability of a given vortex state?,
(ii) why can metastable states exhibit paramagnetism?, and 
(iii) what is the nature of the instability that triggers a change in
the number of vortices?. We anticipate the answers to these questions.
(i) The metastability results  from the sample surface which favors
a large surface superconductivity and opposes both vortex escape and entrance.
This translates into a very stable vorticity or topological charge,
$Q$, associated with all the local minima of the energy functional.
(ii) On decreasing the magnetic field, a reversal in the direction of 
the total supercurrent flow associated with most local minima can take
place before vortices escape. In general, and in contrast with recent work,
detector effects\cite{Peeters:br99} need not be invoked
to explain the paramagnetic response. (iii) The ultimate mechanism
responsible for an instability in the vorticity or change 
in the number of vortices of the system is the disappearance of the saddle
point separating a local minimum from a neighboring one with different $Q$. 
This is called energetic instability. We argue below that,
either increasing or decreasing the magnetic field, there seems to be 
experimental evidence that this energetic instability is preempted by
other mechanism, either associated with noise or some other type
of fluctuations (thermal or quantum).

We start from the Ginzburg-Landau functional for the Gibbs
free energy difference between the normal and superconducting states in
an external magnetic field $H$:
\begin{eqnarray}
G&=&\int d{\bf r} \left[ \alpha |\Psi({\bf r})|^2 + 
\frac{\beta}{2}|\Psi({\bf r})|^4 + \right.\nonumber \\
&&\left.\frac{1}{2m^*}\left|\left(-i\hbar{\bf\nabla} - 
\frac{e^*}{c}{\bf A({\bf r})}\right)\Psi({\bf r})\right|^2 +
\frac{[h({\bf r})-H]^2}{8\pi} \right],
\label{G-L}
\end{eqnarray}
where $\Psi({\bf r})$ is the order parameter or Cooper pair wave
function, $h({\bf r})=\bf\nabla \times 
{\bf A}({\bf r})$, and $\alpha$ and $\beta$ are
the condensation and interaction energy parameters,
respectively. Numerical minimization procedures have been used 
in the past\cite{Peeters:br99,Peeters:prl98,Peeters} to find global and
even local minima of the Ginzburg-Landau
functional applied to mesoscopic superconducting disks. 
However, saddle points or energy barriers, which
are essential for the analysis of the stability of the local minima,
cannot be easily obtained from these methods. Before going into
the details of how to overcome this problem, a few comments are in order.
It is well known that the magnetic response to an external magnetic field  
of a superconductor varies with its size, geometry and orientation with
respect to the field in a non-trivial way. Type-I superconductors expel
the magnetic field below the critical temperature, 
but, for the largest disks in Geim's experiment,
the magnetic field can penetrate the interior\cite{Geim:97,Geim:98}. 
Moreover, multiple-vortex structures\cite{Peeters:prl98,Palacios:disk}, 
only expected in Type-II superconductors,
clearly reflect in the FC measurements\cite{Geim:98} (see below). In
summary, these disks behave like Type-II superconductors, which makes it
possible, in a range of fields, to consider a uniform magnetic 
induction $h({\bf r})=B$\cite{Palacios:disk}.
Next, we project the order parameter onto the "lowest Landau level"
subspace\cite{Palacios:disk,Palacios:strip}:
\begin{equation}
\Psi({\bf r})=\sum_L C_L \frac{1}{\ell\sqrt{2\pi}}
e^{-iL\theta}\Phi_L(r).
\label{expansion}
\end{equation}
This subspace is 
spanned by normalized eigenfunctions of the linearized differential 
Ginzburg-Landau equations where $L$
is the angular momentum ($\ge 0$) and  $\Phi_L(r)$ is the associated nodeless 
function subject to the boundary conditions of zero current through the 
surface.
In this expansion $C_L \equiv |C_L|e^{i\phi_L}$ are complex coefficients
and the radial unit is the magnetic length $\ell=\sqrt{e^*\hbar/cB}$.
We are considering the thickness of the disk to be smaller than
the coherence length so that the system becomes effectively two-dimensional.
This expansion has been shown to give good qualitative as well as quantitative
results for the equilibrium properties at moderately high 
fields\cite{Palacios:disk}.

The central idea in our method for finding generic stationary solutions 
of the Ginzburg-Landau functional is to project the order
parameter onto smaller subspaces spanned by a finite number $N$ of
eigenfunctions, $\{L_1,L_2,\dots,L_N\}$, where $0 \leq L_1<L_2<\dots<L_N$. 
We will restrict our discussion to a disk radius $R= 5\xi(0)$ which
approximately corresponds to the largest disk in Geim's 
experiment\cite{Geim:98} 
[$\xi(0)$ is the coherence length at $T=0$]. For such a disk size
subspaces with dimension $N>3$ do not play any
role\cite{Palacios:disk} and the projected Ginzburg-Landau functional 
reduces to
\begin{eqnarray}
G&=&(B-H)^2+\sum_{i=1}^{N} \alpha[1-B\epsilon_{L_i}(B)]
|C_{L_i}|^2 \nonumber \\ +
&&\frac{1}{4}\alpha^2\kappa^2 B R^2 \times 
\left[\sum_{i=1}^{N} I_{L_i}(B) |C_{L_i}|^4+\right. \nonumber \\
&& \left. \sum_{j>i=1}^N 4 I_{L_iL_j}(B)|C_{L_i}|^2|C_{L_j}|^2 \right]+ ...,
\label{LLL}
\end{eqnarray}
where the Gibbs free energy is expressed
in units of $H_{\rm c2}^2V/8\pi$  ($V$ being the volume
of the disk), $\epsilon_L(B)$ is the energy of the quantum state $L$ 
expressed in units of $\hbar\omega_c/2$ 
($\omega_c=e^*B/m^*c$), $R$ is expressed in units of the coherence length
$\xi(T)$ and $B$ and $H$ are given in units of $H_{\rm c2}(T)$.
The interactions appear in $I_L(B) 
\equiv \int dr\:r\: \Phi_{L}^4$, which can be interpreted as 
the interaction between Cooper pairs occupying the same
quantum state $L$, and $I_{L_iL_j}(B)\equiv
\int dr\:r\: \Phi_{L_i}^2 \Phi_{L_j}^2$,
accounting for the interaction between Cooper pairs occupying different
quantum states. Interaction terms that depend
on the phases of the coefficients appear 
when $L_1+L_3=2L_2$\cite{Palacios:disk}, 
but they are not considered here since these subspaces 
do not play any role in our discussion. 
Stationary solutions of the projected functional (\ref{LLL}) can thus be
found analytically with respect to $|C_L|$ and numerically
with respect to $B$.

 \begin{figure}
\centerline {\epsfxsize=8cm \epsfbox{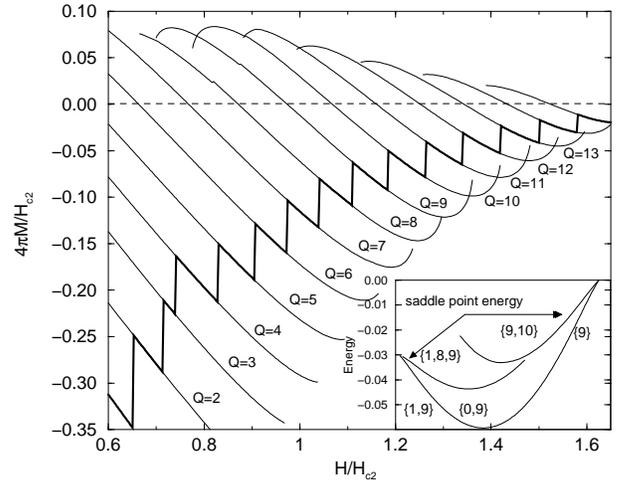}}
\caption{CT magnetization curves associated with local minima of
different topological charges $Q$ for a disk radius $R=5\xi(0)$ and
$\kappa=1$. The equilibrium magnetization is
represented by the thick solid line. Inset: Energy along the local minimum 
$Q=9$ compared to the energy of the saddle point that separates it from the
closest neighboring local minima.}
\label{CT}
\end{figure}
                                                 
\begin{figure}
\centerline{\epsfxsize=5.5cm \epsfbox{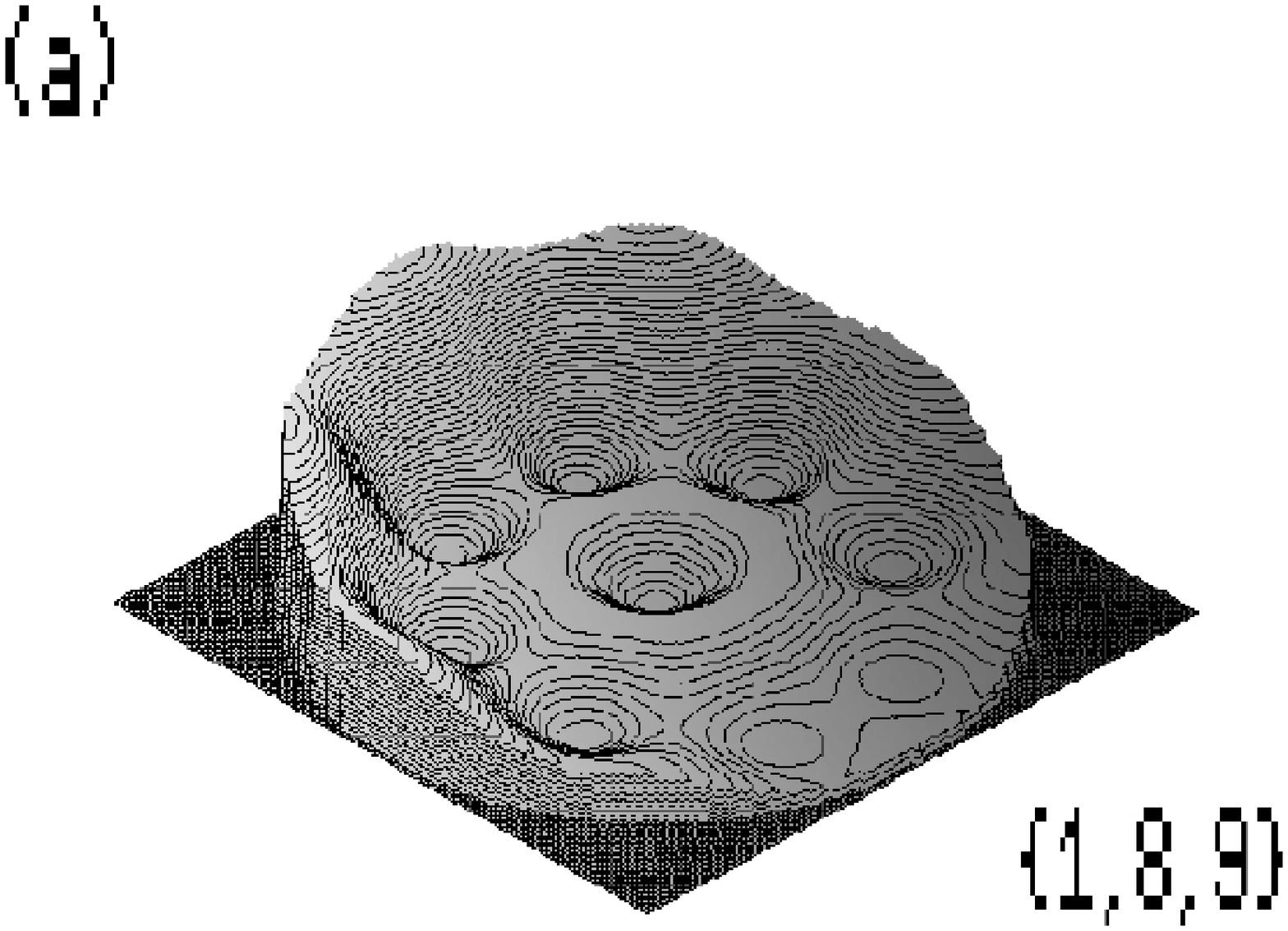}}
\centerline{\epsfxsize=5.5cm \epsfbox{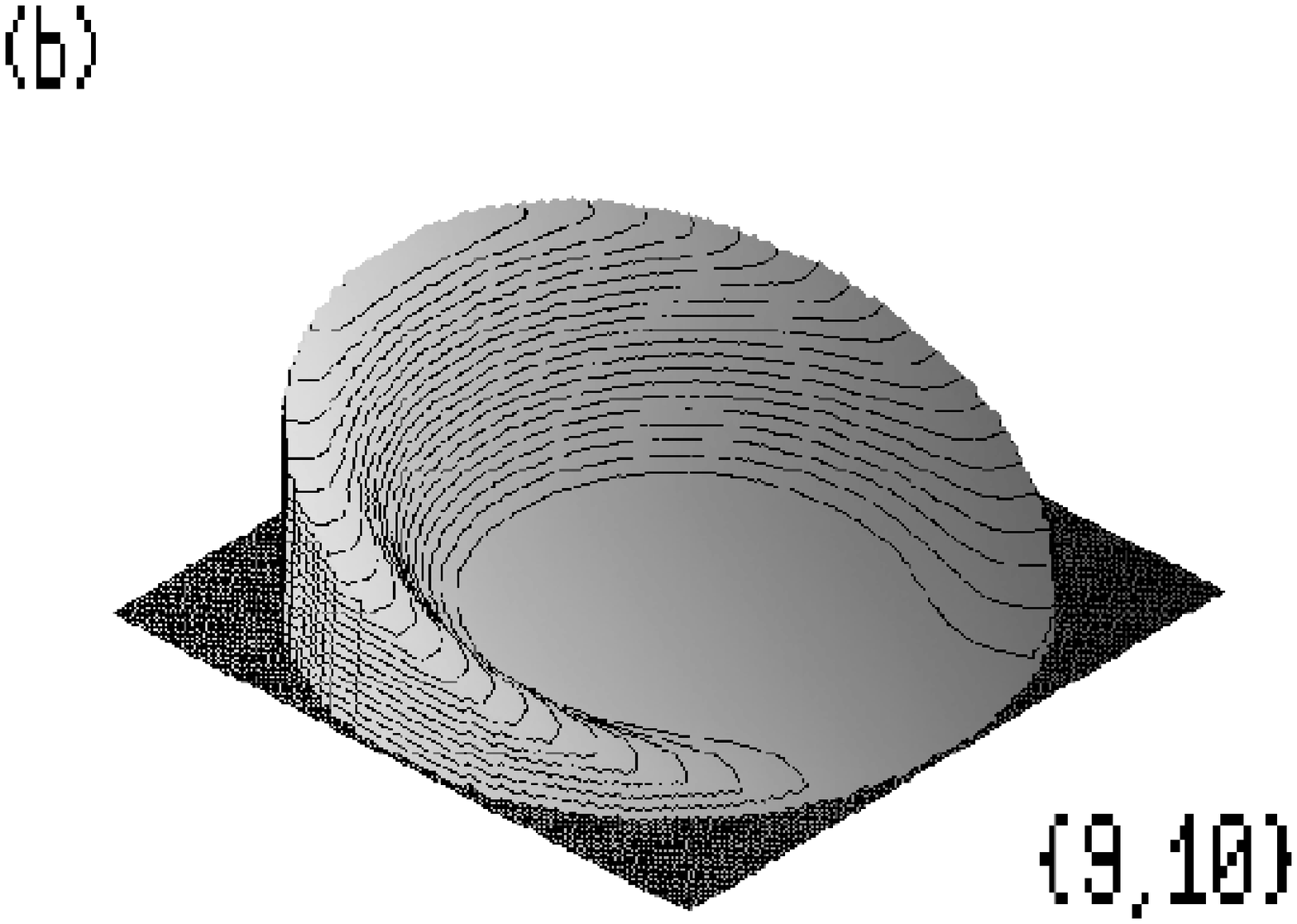}}
\caption{Modulus of the order parameter at the saddle points separating
the local minimum (a) $\{1,9\}$ 
from $\{1,8\}$ close the low-field end of the $Q=9$ curve and (b)
$\{9\}$ from $\{10\}$ close the high-field end of the same curve.}
\label{saddle}
\end{figure}

As we have anticipated, in the expression (\ref{LLL})
$\kappa$ does not take on the bulk nominal
value for Al, but an effective one 
that takes into account the geometry of the disk. 
It is very difficult to estimate this effective value, 
but the experimental evidence of
the existence of multiple-vortex structures for the $R=5\xi(0)$ disk
along with recent detailed numerical calculations\cite{Peeters:prl98}
indicate that $\kappa \gtrsim 1$.  We will consider $\kappa =1 $ throughout.
The main results of this work are shown in Fig. \ref{CT} which shows  
the magnetization associated with all the energetically stable stationary
solutions, i.e., all the local minima.
These local minima are characterized by a
vorticity or topological charge $Q$ which is defined as the number of 
times that the phase of the order parameter winds around $2\pi$ as we complete
one circumference moving along the inner surface of the disk. 
Different curves correspond to different values of $Q$. 
Along any of these curves the topological charge distributes itself
in a variety of ways. For large $Q$ ($Q \geq 12$),
the local minimum is always a giant
vortex with $L=Q$. This local minimum is separated from neighboring
local minima with $L=Q+1$ and $L=Q-1$ by saddle points which appear as 
energetically unstable stationary solutions in the subspaces $\{Q,Q+1\}$ and
$\{Q-1,Q\}$, respectively. Figure \ref{saddle}(b) shows an example of 
the modulus of
the order parameter at the saddle point that separates the giant
vortex $L=9$ from the  $L=10$ close to the high-field end of the $Q=9$ curve. 
Notice the strong depletion of the order parameter at an (arbitrary) point 
on the surface\cite{Peeters:prl99}. 
Higher-energy saddle points also appear separating 
the giant vortex $L=Q$ from other possible local minima $L=Q+2$, 
$L=Q-2$, etc., which fully guarantees its energetic stability.
At the low- and high-field ends of the curve the local minimum $L=Q$
merges with the saddle points $\{Q-1,Q\}$ and
$\{Q,Q+1\}$,  respectively (see, for instance,
 the high-field side of the inset in Fig.
\ref{CT}). In the presence of dissipation
the system will be driven toward the neighboring local minimum. (A dynamical 
analysis\cite{Aranson} of this process would be 
interesting, but it is beyond the scope of this work.) For
smaller $Q$ ($Q < 12$) a giant vortex can be typically found close to the
high-field end of the curve, but, as we move along the curve towards lower 
values of $H$, we cross the critical field where
the multiple vortex solution in the form of a ring, $\{0,Q\}$,
becomes energetically favorable\cite{Peeters,Palacios:disk}.
There is no barrier separating them and a weak second order phase 
transition takes place. 
Similarly, there are saddle points separating the solution 
$\{0,Q\}$ from the local minima $\{0,Q-1\}$ and $\{0,Q+1\}$ on neighboring
curves. These barriers appear as stationary solutions in the subspaces
$\{0,Q-1,Q\}$ and $\{0,Q,Q+1\}$, respectively.
There is also a barrier separating the local minimum $\{0,Q\}$ from
a local minimum $\{1,Q\}$ which may become energetically favorable 
as we move upwards along the curve. There is a saddle point separating
them which appears as a stationary solution in the
subspace $\{0,1,Q\}$. These saddle points are typically one or
two orders of magnitude smaller than those separating states with
different topological charges, although, at the energy crossing, they
still range from $\approx 50$ K ($Q=8$) to $\approx 0.1$ K ($Q=11$).
The experimental temperature is 0.4 K, which means that the system could
evolve along the state $\{0,Q\}$ in the absence
of fluctuations. Otherwise, the system would experience a weak 
first order transition on crossing the 
barrier and changing the vortex structure into a ring with a vortex in
the middle\cite{Geim:note}. 
We have chosen this possibility in our plot, although 
these first order transitions barely reflect in the magnetization.
More structural changes and more weak first order 
transitions can take place as one moves along the curve on decreasing
$H$. At the low-field end of the curve,
the saddle point separating the local minimum $\{L_1,\dots, Q\}$ from 
$\{L_1,\dots, Q-1\}$, i.e., the $\{L_1,\dots,Q-1,Q\}$ stationary solution, 
merges with the local minimum $\{L_1,\dots,Q\}$ (see, for instance,
the low-field side of the inset in Fig. \ref{CT}) and the energetic
instability sets in. Before reaching the low-field end, 
as Fig. \ref{saddle}(a) shows, the
superconducting density exhibits a strong depletion
at some point on the surface when crossing the saddle point.

As Fig. 1 shows, the magnetization associated with the local minima 
changes sign at some intermediate value of $H$ (which depends on $Q$).
In other words, the supercurrent reverses the direction in which it flows. 
At this point the height of the vortex escape
barrier approaches, typically, its maximum value (see inset in Fig. 1
where the zero-magnetization point coincides with the minimum in the 
free energy). As already mentioned,
these barriers can be several orders of magnitude the experimental
temperature. Thus, whatever mechanisms may be operating to decrease 
the topological charge are unlikely
to be efficient enough to prevent the appearance of the paramagnetic 
response on decreasing $H$. In fact, this is what is observed in the 
experiment. The sign change in the response occurs approximately when
the dominant eigenfunction in the expansion of the order parameter,
$L_N=Q$, crosses the minimum of the band structure $\epsilon_L(B)$ and
reverses its group velocity. 
As was shown in Refs. \onlinecite{Palacios:disk,Palacios:strip}, 
the minimum in the band structure comes about due to the boundary
conditions imposed to the components of the order parameter. 
Notice that detector size effects need not be invoked\cite{Peeters:br99} 
to account for the paramagnetic response.

The experimental magnetization curves\cite{Geim:98} 
are, to a large extent, similar to the theoretical ones for small $Q$. 
However, for large $Q$, where our method is expected to be more 
reliable, there are significant
discrepancies. The derivative with respect to the field of the
magnetization curves changes sign close to the
low- and high-field ends of the curves. Furthermore, neighboring curves 
even get to cross each
other. This is never seen in the experiment\cite{Geim:98} which 
seems to indicate that
a vortex can escape or enter the disk before the saddle point or barrier
disappears. Since thermal activation is only efficient
very close to both ends of the curves, either experimental noise or quantum
fluctuations may be ultimately responsible for the vorticity change. 

Even if the energetic instability on the diamagnetic side
is preempted by some of the relaxation processes mentioned above,
the magnetization associated with the global minimum (thick line in Fig.
\ref{CT}) is not likely to be observed for increasing field 
without intentional relaxation. It has been suggested
in the literature that surface roughness is responsible for the
destruction of the saddle points associated to the surface barrier.
The saddle points preventing the escape or entrance of
vortices have the same origin and  surface roughness should affect 
them similarly. Thus, in our view, there are no
reasons for the system to follow the ground state on decreasing $H$ 
and it is expected to continue along a given curve until the
onset of the energetic instability or until vortex 
entrance rates increase to typical measurement times. 

\begin{figure}
\centerline {\epsfxsize=6.0cm \epsfbox{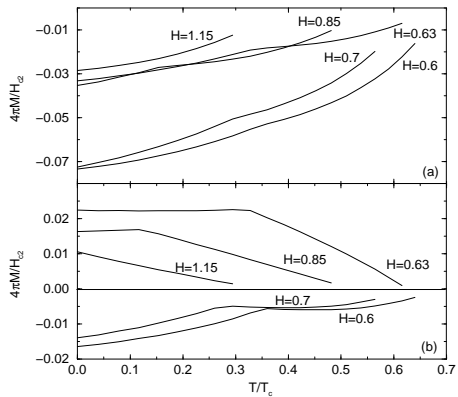}}
\caption{(a) FC magnetization for a disk of radius $R=5\xi(0)$ and
$\kappa=1$ at different values of the external magnetic field. 
Each curve corresponds
to the topological charge $Q_n$ of the giant vortex that nucleates at the
highest critical temperature. (b) The same as in (a), 
but for a topological charge  $Q_n+1$.}
\label{FC}
\end{figure}

Finally, we would like to comment on the FC measurements.  The current
understanding of the FC results is summarized in a work by Moshchalkov et
al.\cite{Moshchalkov:prb97} which attributes the FC paramagnetic response to a
flux-compression phenomenon.
Figure \ref{FC} shows the magnetization as a function of temperature for
different values of the magnetic field. The usual phenomenological 
temperature scaling of the parameters in the Ginzburg-Landau
functional (\ref{G-L}) has been considered\cite{Moshchalkov:prb97}. 
Each curve in Fig. \ref{FC}(a) corresponds
to the topological charge $Q_n$ of the giant vortex that nucleates at the
highest possible critical temperature for each chosen field. 
$Q_n$ is maintained along the
different curves down to $T=0$ due to presence of the energy barriers
discussed above which prevent the change of vorticity. 
The response is always diamagnetic in clear contrast with the FC 
data. Figure \ref{FC}(b) shows magnetization curves
for a topological charge  $Q_n+1$.
Alternating  paramagnetic and diamagnetic behaviors are obtained as a
function of $H$ and a low-temperature saturation of the paramagnetic 
response occurs due to explosion of the giant vortex as suggested by
Moshchalkov et al.\cite{Moshchalkov:prb97}. This behavior is in remarkable
agreement with the FC data\cite{Geim:98} which seems to
suggest that either thermal fluctuations close to the
critical temperature or surface roughness favor the nucleation
of giant vortices with a higher topological charge than that expected
from plain Ginzburg-Landau theory\cite{note}. 
Notice again that, in our approach, the magnetic field is uniform in space
which suggests that flux compression\cite{Geim:98,Moshchalkov:prb97} 
is not essential as far as the existence of paramagnetism is concerned.
Still, the origin of the PME in the FC measurements remains an open issue.

In closing, we have addressed some fundamental questions posed by the
experiment of Geim et al.\cite{Geim:98}. Saddle points or barriers of
the Ginzburg-Landau functional have been found based on a projection
technique. This has allowed us to obtain metastable magnetization
curves and to gain insight into the controversial paramagnetic response
both in FC and CT magnetization measurements.

The author acknowledges enlightening discussions with J. Dukelsky,
J. Fern\'andez-Rossier, J. J. Garc\'{\i}a-Ripoll, A. K. Geim, F. Guinea,
B. Paredes, G. Sierra, and C. Tejedor. This
work has been funded by MEC of Spain under contract No. PB96-0085.


\begin{references}
\bibitem{note}Similar difficulties were found in Ref.[14] using a numerical
method that considers a non-uniform magnetic induction
$b(r)$. The authors, however, attributed the discrepancy between theory and
experiment to numerical uncertainty in their calculations.
\end{references}
\end{document}